\documentclass[fleqn,twoside]{article}
\usepackage{espcrc2}

\usepackage{graphicx}

\title{
\vspace{-0.7cm}
\hfill \rm \null \hfill
 \hbox{\normalsize ADP-02-86/T525} \\
\vspace{-2mm}
 \hfill \hbox{\normalsize JLAB-THY-02-45} \\
\vspace{-1mm}
Excited baryons from the FLIC fermion action}

\author{
W.~Melnitchouk\address[CSSM]{Special Research Center for the
	Subatomic Structure of Matter, and		\\
	Department of Physics and Mathematical Physics,
	University of Adelaide, 5005, Australia}$^,$
        \address[JLab]{Jefferson Lab, 12000
	Jefferson Avenue, Newport News, VA 23606, U.S.A.}
\thanks{Presented by W.~Melnitchouk},
J.~N.~Hedditch\addressmark[CSSM],
D.~B.~Leinweber\addressmark[CSSM],
A.~G.~Williams\addressmark[CSSM], 
J.~M.~Zanotti\addressmark[CSSM] and
J.~B.~Zhang\addressmark[CSSM]
}
       
\begin{document}

\begin{abstract}
Masses of positive and negative parity excited nucleons and hyperons
are calculated in quenched lattice QCD using an ${\cal O}(a^2)$
improved gluon action and a fat-link clover fermion action in which
only the irrelevant operators are constructed with fat links.
The results are in agreement with earlier $N^*$ simulations with
improved actions, and exhibit a clear mass splitting between the
nucleon and its parity partner, as well as a small mass splitting
between the two low-lying $J^P={1\over 2}^-$ $N^*$ states.
Study of different $\Lambda$ interpolating fields suggests a similar
splitting between the lowest two ${1\over 2}^-$ $\Lambda^*$ states,
although the empirical mass suppression of the $\Lambda^*(1405)$ is
not seen.
\end{abstract}

\maketitle

\section{INTRODUCTION}

The study of baryon excitations provides valuable insight into the
forces of confinement and the nature of QCD in the nonperturbative regime.
This is one of the motivations for the experimental effort currently
under way at Jefferson Lab which is accumulating data of unprecedented
quality and quantity on $N \to N^*$ transitions.

One of the long-standing puzzles in baryon spectroscopy
has been the low mass of the first positive parity
excitation of the nucleon (the $N^*(1440)$ Roper resonance) compared
with the lightest odd parity state.
Without fine tuning of parameters, valence quark models tend to leave 
the mass of the Roper too high.
%
Another challenge is presented by the odd parity $\Lambda^*(1405)$,
whose anomalously small mass has been interpreted as an indication of
strong coupled channel effects, and a weak overlap with a three-valence
quark state.
While model studies have provided some understanding of the level
orderings and mass splittings in the baryons spectrum, it is
hoped that lattice QCD \cite{LEIN1,DEREK,RICHARDS,DWF} will provide a
definitive resolution of some of the outstanding issues.

%
Recently a new approach to nonperturbative ${\cal O}(a)$ fermion
action improvement has been developed based on the Fat-Link Irrelevant
Clover (FLIC) fermion action \cite{FATJAMES}.  First results for the
spin-1/2 excited nucleon and hyperon, and spin-3/2 $N^*$ and
$\Delta^*$, spectra were presented in Refs.~\cite{NSTAR} and
\cite{SPIN32}, respectively.
Here we update the earlier analyses \cite{NSTAR} by including more
configurations, and using correlation matrix techniques to better
resolve individual excited states of the nucleon and $\Lambda$.

\section{INTERPOLATING FIELDS}

Following standard procedure, we calculate two-point correlation
functions from baryon interpolating fields constructed to maximize
overlap with states with specific quantum numbers.
In this analysis two interpolating fields are considered.
For the positive parity proton these are given by,
\begin{eqnarray}
\label{chi1p}
\chi_1^{p +}(x)
&=& \epsilon_{abc}
\left( u^T_a(x)\ C \gamma_5\ d_b(x) \right) u_c(x)\ ,	\\
\label{chi2p}
\chi_2^{p +}(x)
&=& \epsilon_{abc}
\left( u^T_a(x)\ C\ d_b(x) \right) \gamma_5\ u_c(x)\ ,
\end{eqnarray}
where the fields $u$, $d$ are evaluated at Euclidean space-time point
$x$, and $a, b$ and $c$ are color labels.

As explained in Refs.~\cite{LEIN1,NSTAR}, because of the Dirac
structure of the ``diquark'' in the parentheses in
Eqs.~(\ref{chi1p}) and (\ref{chi2p}), one expects correlation functions
constructed from $\chi_2^{p+}$ to be dominated by larger mass states
than those arising from $\chi_1^{p+}$.
Moreover, 
%
the $\chi_2^{p+}$ interpolating field
%
%
is known to have little overlap with the nucleon ground state
\cite{LEIN1,BOWLER}.

For the $\Lambda$ states, we consider several interpolating fields
with different SU(3) properties.
To test the extent to which SU(3) flavor symmetry is valid in
the baryon spectrum, we construct an interpolating field
(``$\Lambda^c$'') composed of terms common to both the singlet
and octet $\Lambda$ interpolating fields \cite{LDW},
\begin{eqnarray}
\label{chi1lc}
\chi_1^{\Lambda^c}(x)
&=& {1\over\sqrt{2}} \epsilon_{abc}
\left\{
   \left( u^T_a(x)\ C \gamma_5\ s_b(x) \right) d_c(x)\
\right.		\nonumber\\
& &
\hspace*{0.3cm}
\left.
  -\ \left( d^T_a(x)\ C \gamma_5\ s_b(x) \right) u_c(x)
\right\} ,
\end{eqnarray}
and similarly for $\chi_2^{\Lambda^c}$.
Such interpolating fields allow for mixing between singlet and octet
states induced by SU(3) flavor symmetry breaking, and may be useful
in determining the nature of the $\Lambda^*(1405)$ resonance.

\section{RESULTS \& DISCUSSION}

The simulations are performed on a $16^3\times 32$ lattice at
$\beta=4.60$, with a lattice spacing of $a = 0.122(2)$~fm, based on a
sample of 392 configurations.  For the gauge fields, a mean-field
improved plaquette plus rectangle action is used, while for the quark
fields, the FLIC action \cite{FATJAMES} is implemented with $n=4$
sweeps of APE smearing at $\alpha=0.7$.  Further details of the
simulations are given in Ref.~\cite{FATJAMES}.

\begin{figure}[t]
\begin{center}
{\includegraphics[width=\hsize]{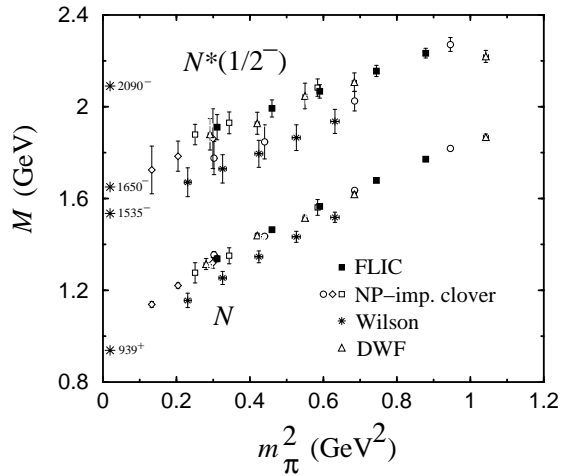}}
\vspace*{-1cm}
\caption{Masses of the nucleon ($N$) and the lowest $J^P={1\over 2}^-$
        excitation (``$N^*$''), obtained from the $\chi_1$ interpolating
        field.  The FLIC and Wilson results are from the present
        analysis.}
\end{center}
\vspace*{-1cm}
\end{figure}

In Fig.~1 we show the $N$ and $N^*({1\over 2}^-)$ masses as a
function of $m_\pi^2$ for the new simulations with the FLIC action.
For comparison, we also show results from earlier simulations with
Wilson \cite{NSTAR} and domain wall fermions (DWF) \cite{DWF},
and a nonperturbatively improved clover action 
\cite{RICHARDS} with different source smearing and volumes.

There is excellent agreement between the different improved actions
for the nucleon mass.
The Wilson results lie systematically low in comparison to these
due to large ${\cal O} (a)$ errors in this action \cite{FATJAMES}.
A similar pattern is repeated for the $N^*({1\over 2}^-)$ masses.
A mass splitting of around 400~MeV is clearly visible between the $N$
and $N^*$ for all actions, including the Wilson.
The trend of the $N^*({1\over 2}^-)$ data with decreasing $m_\pi$
is also consistent with the mass of the lowest physical negative
parity $N^*$ state.

\begin{figure}[t]
\begin{center}
{\includegraphics[width=\hsize]{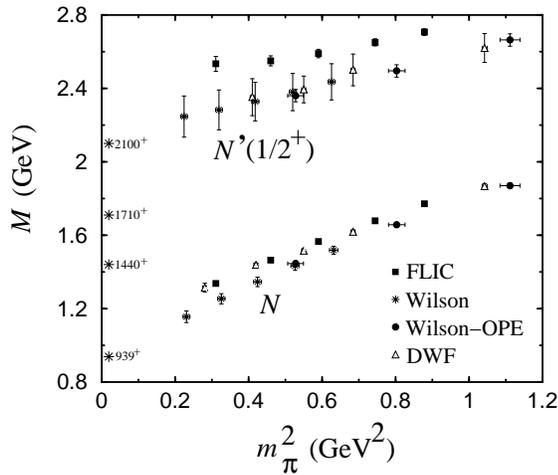}}
\vspace*{-1cm}
\caption{Masses of the nucleon, and the lowest $J^P={1\over 2}^+$
        excitation (``$N'$'') obtained from the $\chi_2$ interpolating
        field.  The FLIC and Wilson results are from this analysis.}
\end{center}
\vspace*{-1cm}
\end{figure}

The mass of the $J^P = {1\over 2}^+$ state obtained from the $\chi_2^{p^+}$
interpolating field is shown in Fig.~2.
%
%
In addition to the FLIC and Wilson results from the present analysis,
also shown are the DWF results \cite{DWF}, and results from an
earlier analysis with Wilson fermions together with the operator
product expansion \cite{LEIN1}.
The most striking feature of the data is the relatively large
excitation energy of the $N'$, some 1~GeV above the nucleon.
It has been speculated that the $\chi_2^{p^+}$ field may have overlap
with the lowest ${1\over 2}^+$ excited state \cite{DWF}, however,
there is little evidence that this state is the $N^*(1440)$.
%
%
It is likely that the $\chi_2$ interpolating field simply does not
have good overlap with either the nucleon or the $N^*(1440)$,
but rather a (combination of) excited ${1\over 2}^+$ state(s).


\begin{figure}[t]
\begin{center}
{\includegraphics[width=\hsize]{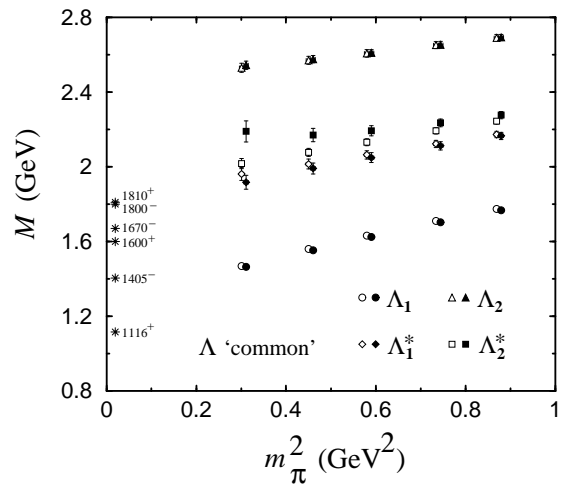}}
\vspace*{-1cm}
\caption{Masses of the $\Lambda({1\over 2}^\pm)$ states, obtained from
	the $\chi^{\Lambda^c}_1$ and $\chi^{\Lambda^c}_2$ interpolating
	fields.  The symbols are described in the text.}
\end{center}
\vspace*{-1cm}
\end{figure}

The spectrum of positive and negative parity $\Lambda$ states
is shown in Fig.~3.
The positive (negative) parity states labeled $\Lambda_1$
($\Lambda_1^*$) and $\Lambda_2$ ($\Lambda_2^*$) are constructed from
the $\chi_1^{\Lambda^c}$ and $\chi_2^{\Lambda^c}$ interpolating
fields, respectively.
The pattern of mass splittings is similar
to that observed for the $N^*$'s in Figs.~1 and 2.
The importance of the correlator matrix analysis (filled symbols) is
evident from a comparison with the naive results (open symbols),
where the states have not been diagonalized.
In particular, using the naive fitting scheme, it is difficult to
obtain a mass splitting between $\Lambda_1^*$ and $\Lambda_2^*$.
Once the correlation matrix analysis is performed it is
possible to resolve two separate mass states.
%
There is little evidence that the $\Lambda_2$ has any
significant overlap with the first positive parity excited state,
$\Lambda^*(1600)$.

While it seems plausible that nonanalyticities in a chiral
extrapolation \cite{MASSEXTR} of $N_1$ and $N_1^*$ results could
eventually lead to agreement with experiment, the situation for the
$\Lambda^*(1405)$ is not as compelling.
Whereas a 150~MeV pion-induced self energy is required for the
$N_1 ,\ N_1^*$ and $\Lambda_1$ states, 400~MeV is required to
approach the empirical mass of the $\Lambda^*(1405)$.
This large discrepancy suggests that relevant physics may be absent
from simulations in the quenched approximation.
%
%
The study of more exotic interpolating fields may
indicate the the $\Lambda^*(1405)$ does not couple strongly to
$\chi_1^\Lambda$ or $\chi_2^\Lambda$.
Investigations at lighter quark masses involving quenched chiral
perturbation theory will assist in resolving these issues.

%

\vspace*{0.3cm}

This work was supported by the Australian Research Council.
W.M. is supported by the U.S. Department of Energy contract
\mbox{DE-AC05-84ER40150},
under which the Southeastern Universities
Research Association (SURA) operates the Thomas Jefferson National
Accelerator Facility (Jefferson Lab).



\begin{thebibliography}{99}

\bibitem{LEIN1}
D.~B.~Leinweber,
Phys. Rev. D {\bf 51}, 6383 (1995).

\bibitem{DEREK}
F.~X.~Lee and D.~B.~Leinweber,
Nucl. Phys. Proc. Suppl. {\bf 73}, 258 (1999).
%

\bibitem{RICHARDS}
C.~R.~Allton {\it et al.},
Phys. Rev. D {\bf 47}, 5128 (1993);
%
%
D.~G.~Richards {\em et al.},
Nucl. Phys. Proc. Suppl. {\bf 109}, 89 (2002);
%
M.~G\"ockeler {\em et al.},
Phys. Lett. B {\bf 532}, 63 (2002).

\bibitem{DWF}
S.~Sasaki, T.~Blum and S.~Ohta,
Phys. Rev. D {\bf 65}, 074503 (2002).
%

\bibitem{FATJAMES}
J.~M.~Zanotti {\it et al.}, 
Phys. Rev. D {\bf 60}, 074507 (2002);
Nucl. Phys. Proc. Suppl. {\bf 109}, 101 (2002).

\bibitem{NSTAR}
W.~Melnitchouk {\em et al.},
hep-lat/0202022;
%
Nucl. Phys. Proc. Suppl. {\bf 109}, 96 (2002).

\bibitem{SPIN32}
J.~M.~Zanotti {\it et al.},
these proceedings.

\bibitem{BOWLER}
K.~Bowler {\it et al.},
Nucl. Phys. {\bf B240}, 213 (1984).

\bibitem{LDW}
D.~B.~Leinweber {\em et al.},
Phys. Rev. D {\bf 46}, 3067 (1992);
Phys. Rev. D {\bf 43}, 1659 (1991).

\bibitem{MASSEXTR}
D.~B.~Leinweber {\em et al.},
Phys. Rev. D {\bf 61}, 074502 (2000).
 


\end{thebibliography}
\end{document}